\documentclass[11pt]{amsart}

\usepackage{amsthm}
\usepackage{amsmath}
\usepackage{amssymb}
\usepackage{a4wide}
\usepackage{bbm}
\usepackage{graphicx}
\usepackage{changepage}
\usepackage{color}                    

\newcommand{\R}{\mathbb{R}}

\newcommand{\set}[1]{\left\{#1\right\}}

\begin{document}

\title{Uncovering Proximity of Chromosome Territories using Classical Algebraic Statistics}
\author{Javier Arsuaga}
\address{Department of Mathematics, San Francisco State University, San Francisco CA 94132}

\author{Ido Heskia}

\author{Serkan Ho\c{s}ten}

\author{Tatsiana Maskalevich}


\begin{abstract}
Exchange type chromosome aberrations (ETCAs) are rearrangements of the genome that occur when chromosomes break and the resulting fragments rejoin with other fragments from other chromosomes. 
ETCAs are commonly observed in cancer cells and in cells exposed to radiation. The frequency of these chromosome rearrangements is correlated with their spatial 
proximity, therefore it can be used to infer the three dimensional organization of the genome. Extracting statistical significance of spatial proximity from cancer 
and radiation data  has remained somewhat elusive because of the sparsity of the data. We here propose a
new approach to study the three dimensional organization of the genome using algebraic statistics. We test our method on a published data set of irradiated human blood 
lymphocyte cells. 
We provide a rigorous method for testing the overall organization of the genome, and in agreement with previous results we find a random relative positioning of 
chromosomes with the exception of  the chromosome  pairs \{1,22\} and \{13,14\} that have a significantly larger number of ETCAs than the rest of the chromosome pairs 
suggesting their spatial proximity. We conclude that algebraic methods can successfully be used to analyze genetic data and have potential applications to larger and more complex data sets.
\end{abstract}



\maketitle

\section{Introduction}

\begin{figure}[htbp]
\centering
\makebox{\includegraphics[height=1.75in]{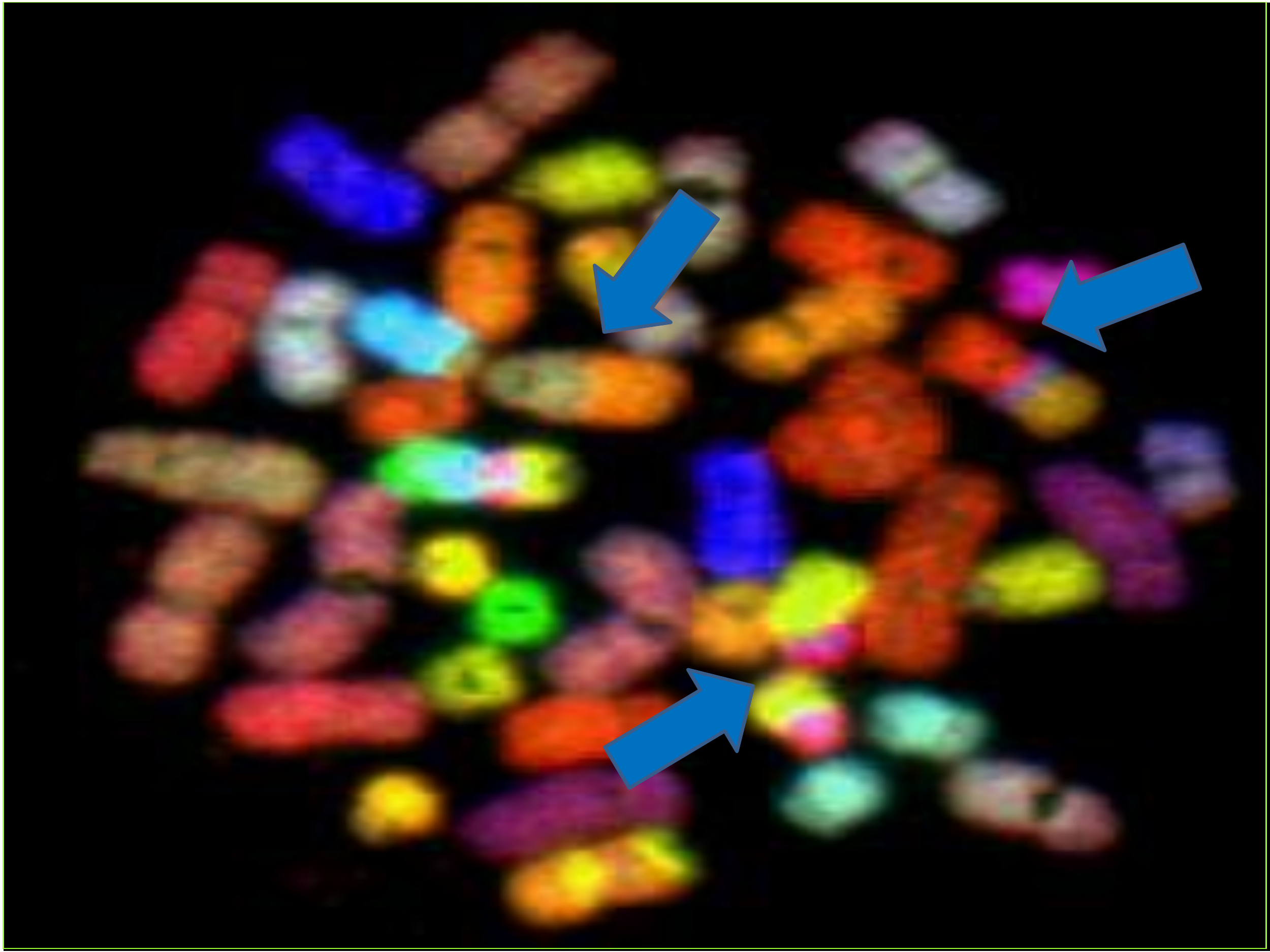}}
\caption{\label{SKY_Image} Examples of ETCAs found in a tumor cell. Pairs of homologue chromosomes were painted the same color using SKY. The arrows point to ETCAs. 
Figure kindly provided by J.L. Garc\'{i}a.}
\end{figure}

\medskip
During the early stages of the cell cycle the mammalian genome is organized in chromosome territories \cite{Cremer82,Rabl} (for a review see \cite{Cremer10}). When DNA damaging agents, such as radiation, 
cross the cell nucleus they introduce double strand breaks that produce chromosome fragments. These chromosome fragments need to be rejoined with their original 
partners for the cell to survive. A small percentage of breaks however are incorrectly rejoined introducing exchange type chromosome aberrations (ETCAs). 
ETCAs between non-homologue chromosomes can be detected in the laboratory by means of diverse chromosome painting and sequencing techniques (see Figure \ref{SKY_Image} for an example of ETCAs detected by the chromosome painting technique SKY). It is expected that chromosomes that are in close spatial 
proximity form ETCAs more often than those that are far apart \cite{Chen96,Bickmore02,Nikiforova,Roix03,ZhangDekker12}. 
Therefore the frequency of ETCAs between non-homologous chromosomes is informative of their relative position and it
can be used to reconstruct the three dimensional structure of the genome.  

Chromosome painting techniques such Fluorescence in-situ Hybridization  (FISH) or its multicolor versions   such as Spectral Karyotyping (SKY) and multiplex FISH (mFISH) 
\cite{Speicher,Schrock} paint every pair of homologue chromosomes the same color (Figure \ref{SKY_Image}). These techniques have shown that the position of 
chromosome territories with respect to the center of the nucleus is driven by gene density, chromosome size and/or local chromatin geometry \cite{Boyle2001,Kolbl,Parada}.
In human lymphocytes gene-rich chromosomes  such as  \{1, 19, 17, 22\}  are located near the center of the nucleus \cite{Boyle2001,Cremer10} while gene-poor chromosomes such as $\{2,4,13,18\}$ are located closer to the periphery of the nucleus \cite{Boyle2001, Cremer10}. 
Studies measuring the relative position of chromosomes, using radiation induced ETCAs, on the other hand have shown that chromosomes are randomly located with respect to each other with the exception of a few chromosome clusters \cite{Corn2002,Hlatky2002,Sachs1997,Sachs2004}. Interestingly  this overall random relative organization has been corroborated by new sequencing techniques \cite{Lieberman09,ZhangDekker12}. 
\medskip

Quantification of the relative position of chromosome territories commonly use tables whose entries are the number of ETCAs detected for any two non-homologue chromosomes 
such as the one we use in  Table \ref{table:Lympho2004}.  Several methods have been proposed to study frequency tables of radiation induced ETCAs. 
In \cite{Boei06} tables of radiation induced ETCAs detected by FISH were analyzed. 
In this type of studies frequency tables were densely populated and with large entries, therefore chi-square statistics were used to find significant clusters of chromosomes.  
In \cite{Ars2004,Corn2002} similar tables were generated employing mFISH data. Although more accurate in some respects, the latter data were more sparse and with small entries. 
In this case the proximity of chromosome territories was tested by assigning p-values to clusters of previously reported  chromosome territories. 
Here we  propose a model-based approach that builds a simple log-linear model to test the proximity of pairs of chromosome territories, 
referred to as chromosome pairs from now on,  and we use a Markov Chain Monte Carlo method based on the theory of algebraic statistics to assign significance.  The mFISH data we analyze have already appeared in the literature \cite{Ars2004,Corn2002,Con2001}.  
\medskip

In our study we test a no-proximity effect model and a single pair proximity model by sampling tables that have the same sufficient statistic as those observed experimentally. 
Sampling of the tables is performed by running a Markov Chain Monte Carlo algorithm that uses a Markov basis \cite{Sturmfels98, LAC} for a second hypersimplex \cite{DST}. 
This kind of method is one of the early contributions of algebraic statistics, hence ``classical algebraic statistics'' in the title of this work.
Our results indicate that we could not reject the hypothesis of random relative arrangement of chromosome territories when radiation induced chromosome aberrations were analyzed. This result is in qualitative agreement with previous studies 
\cite{Ars2004, Corn2002, Con2001} and suggests that the specific positions of any two pairs of chromosomes do not influence the frequency of aberrations observed. 
However, by assuming the existence of a proximity biasing factor we found chromosome pairs $\{1,22\}$ and $\{13,14\}$ to be significant. We conclude that methods
 develop in algebraic statistics are suitable for analyzing genetic data of moderate size in which data sparsity or low numbers of measurements are present.  
 
\section{Data and Methods}

To test for proximity of chromosome territories we used a radiation induced exchange type chromosome aberration table published in  \cite{Ars2004,Vives05}.
 In these experiments cells from healthy donors were irradiated with sparsely ionizing $\gamma$-rays at different doses, and mFISH \cite{Speicher} was used to detect ETCAs. 
This table includes a total of 3585 records of human peripheral blood lymphocytes irradiated with sparsely ionizing radiation at different doses.
\medskip

The frequency of ETCAs was summarized by recording the number of cells in which at least one exchange between two non-homologous chromosomes occurred. This 
quantity is robust with respect to noise introduced by apparently incomplete aberrations (i.e. those aberrations with not all fragments accounted for) and reduces false positives. Following previous publications \cite{Ars2004,Corn2002}
we denote these values by $f(j,k)$ where $j$ and  $k$ are the chromosomes that participate in the exchange. These values are presented in Table \ref{table:Lympho2004}  as a $22\times 22 $ upper-triangular table.

\begin{table}[h]
	\begin{center}
\caption{Table of ETCAs in 3585  human lymphocytes as reported in \cite{Ars2004,Vives05}. Each entry $f(j,k)$ holds the number of cells in which at least one exchange between chromosomes $j$ and $k$ was recorded.  
The total number of cells in which a given chromosome was involved in at least one exchange appears in the "sum" column.}
	{\tiny\tt
	\begin{tabular}{lllllllllllllllllllllll}

Chr&2&3&4&5&6&7&8&9&10&11&12&13&14&15&16&17&18&19&20&21&22&Sum\\
\hline1&44&38&42&29&26&29&18&39&29&25&18&15&18&34&31&22&12&14&22&9&27&541\\
2&&43&37&32&30&24&25&29&16&24&30&29&9&26&8&24&8&7&12&13&15&485\\
3&&&21&31&32&24&21&26&23&25&23&21&18&18&19&21&11&17&11&12&10&465\\
4&&&&23&27&28&24&26&20&13&19&23&22&20&16&18&11&6&12&10&7&425\\
5&&&&&17&31&26&25&24&30&25&25&15&19&8&19&13&7&16&7&4&426\\
6&&&&&&18&22&21&31&13&30&18&15&19&14&15&13&10&9&8&7&395\\
7&&&&&&&20&20&17&28&25&13&18&8&18&23&11&9&19&6&7&396\\
8&&&&&&&&13&12&24&11&25&15&16&12&16&17&4&9&7&8&345\\
9&&&&&&&&&21&25&7&23&23&27&20&15&22&8&9&7&10&416\\
10&&&&&&&&&&18&21&14&14&10&19&14&9&5&11&7&3&338\\
11&&&&&&&&&&&25&5&15&16&19&15&8&10&12&3&11&364\\
12&&&&&&&&&&&&9&16&9&12&16&8&13&10&5&5&337\\
13&&&&&&&&&&&&&29&10&10&7&16&5&6&7&9&319\\
14&&&&&&&&&&&&&&22&13&6&10&2&6&13&11&310\\
15&&&&&&&&&&&&&&&22&13&9&7&11&7&9&332\\
16&&&&&&&&&&&&&&&&12&15&12&20&8&13&321\\
17&&&&&&&&&&&&&&&&&5&4&11&5&10&291\\
18&&&&&&&&&&&&&&&&&&2&11&9&3&223\\
19&&&&&&&&&&&&&&&&&&&6&0&8&156\\
20&&&&&&&&&&&&&&&&&&&&7&10&240\\
21&&&&&&&&&&&&&&&&&&&&&6&156\\
22&&&&&&&&&&&&&&&&&&&&&&193\\
		\end{tabular}}  
	\end{center}
 	
 	 \label{table:Lympho2004}\
\end{table}

\subsection{The No-proximity and Single-pair Proximity Effects as Log-Linear Models} 
ETCAs between two non-homologous chromosomes $j$ and $k$  ($j\ne k$)  were modelled by a single discrete random variable $X$ with 
${22 \choose 2}=231$ values corresponding to all possible pairs of non-homologue chromosomes with a probability density function given by 
$$p \, : \, \{ (j,k) \, : 1 \leq j < k \leq 22 \} \longrightarrow \Delta_{231}$$ where the set  
$$\Delta_{231}  \, = \, \{ (p_{12}, p_{13}, \ldots, p_{2122}) \in \mathbb{R}^{231} \, : \, p_{jk} \geq 0 \,\,\, \sum_{j,k} p_{jk} = 1 \}$$
is the probability simplex in $\mathbb{R}^{231}$ and  $p_{jk} = p(j,k)$. In this no-proximity effect model the probability of observing an ETCA 
between the chromosomes $j$ and $k$ is
$$p_{jk} \, = \, \theta_j \theta_k \,\,\, 1 \leq j < k \leq 22.$$
where $\theta_1, \theta_2, \ldots, \theta_{22}$ are positive parameters. More precisely,  the no-proximity effect model is the image of $\R_+^{22}$ in $\Delta_{231}$ under the map
$\phi(\theta_1, \ldots, \theta_{22}) = (\theta_j \theta_k \, : \, 1 \leq j < k \leq 22)$. This model can be linearized by applying logarithms to both sides of the equation, 
which gives $\log(p_{jk}) = \log(\theta_j) + \log(\theta_k)=\beta_j+\beta_k$ and shows that our model is a log-linear model \cite{LLM}. Therefore the no-proximity effect model is given 
by a $22 \times 231$ design matrix, denoted by $A(22)$, whose columns are $e_j + e_k$  with $1 \leq j < k \leq 22$ where $e_j$ is the $j$th standard unit vector in $\R^{22}$. 
The columns of $A(22)$ are the vertices of the {\it second hypersymplex} in $\R^{22}$~\cite{DST}.

\medskip
For a fixed pair $\{r,s\}$ with $r\ne s$  we define an extended model that we call a {\it single-pair proximity effect model}. This model is given by the map
$\phi': \R_+^{23}  \to \Delta_{231}$
defined as 
\begin{equation}
p_{jk} = \phi'(\theta_1,..,\theta_{22}, \mu_{rs})= \left\{ \begin{array}{ll} \theta_r \theta_s \mu_{rs} &\mbox{ if } j=r \mbox{ and } k=s,\\
   \theta_j \theta_k &\mbox{otherwise.}\end{array}\right.
\end{equation}
The parameter $\mu_{rs}$ is a bias factor for the frequency of an observed exchange between chromosomes $r$ and $s$. This bias factor corresponds to a proximity factor between 
the territories of two chromosomes $r$ and $s$. 
By taking the logarithm on both sides of the above equation we get the parametrization in logarithmic coordinates: 
 	\begin{equation}
	 \log(p_{jk})=
	\left\{ \begin{array}{ll} \beta_r+\beta_s+\alpha_{rs}\;\;\;\; &\mbox{ if } j=r, k=s, \\
	\beta_j+\beta_k  & \hfill \mbox{ otherwise. }
         \end{array}\right. \end{equation}
 
The single-pair proximity effect model for any pair of chromosomes $\{r,s\}$ is also a log-linear model which extends the no-proximity effect model since 
the set of probability distributions in this model are  
those in the image of the map $\phi'$ with $\alpha_{rs}=0$. The design matrix $A[r,s]$ defining this model is a $23 \times 231$ matrix, identical to $A(22)$ in its first $22$ rows, and with an extra row of all zeros except a $1$
in the column corresponding to $p_{rs}$.

\medskip
In order to assign a $p$-value to the goodness-of-fit test we propose to sample tables that are similar to those observed experimentally. More specifically we sample tables with the same {\it minimal sufficient statistic} as
the data table $f$. For each chromosome  $k$ we have the {\em marginal total} $$ u_k  \, = \, \sum_{j: \, j\ne k} f(j,k) \quad \quad  \mbox{for     } k = 1, \ldots, 22.$$ This quantity is displayed in the {\tt Sum} column of Table \ref{table:Lympho2004}. 
The marginal total of a given chromosome is a measure of the propensity of each individual chromosome to form ETCAs. In radiation studies 
this quantity has been associated to the sensitivity of the chromosomes to radiation \cite{Ars2004,Corn2002} and to repair mechanisms \cite{Wu01}. 
 For the no-proximity effect model $\phi$ the collection of marginal totals $u = (u_k: \, k=1, \ldots, 22)$  
is the minimal sufficient statistic. For the single-pair proximity effect model for the chromosome pair $\{r, s\}$, the minimal sufficient statistic is the same marginal sums together with $u_{rs} = f(r,s)$. 
The set of tables with the same sufficient statistic  is called {\em the fiber} of the experimentally observed table and is denoted by $\mathcal{F}(u)$. It is well-known that $\mathcal{F}(u)$ consists of lattice points in a polytope.
 
\subsection{Maximum Likelihood Estimation}

The maximum likelihood estimator (MLE) tables of the data with respect to the no-proximity effect and single-pair proximity models were computed by the standard numerical algorithm  {\it Iterative Proportional Scaling} \cite{IPS}. For log-linear models,
the algorithm converges to the unique MLE table $\hat f$ such that $\frac{1}{N} \hat f$ lies on the model where $N$ is the sample size and $\hat f$ has the same sufficient statistic $u$ as the data table $f$.
The existence of this unique table is guaranteed by Birch's Theorem (see \cite{LAC}). 

\subsection{Hypothesis testing and Monte-Carlo simulations}
Our goodness-of-fit test for the no-proximity effect model  uses the standard chi-square statistic  

\begin{equation} \label{chi-square}
 \chi^2(F)=\sum_{1 \leq j < k \leq 22}  \frac{(F(j,k)-\hat f(j,k))^2}{\hat f(j,k)}.
\end{equation}
where $\hat f(j,k)$ are the entries of the MLE table with respect to the no-proximity effect model
given the data table $f$, and $F(j,k)$ are the entries of tables $F$ drawn from the fiber $\mathcal{F}(u)$.
Typically large $\chi^2(F)$ values  would indicate that the data table $f$ is ``close'' to $\hat f$ providing no evidence for rejecting no-proximity effect model.  
\medskip


The fiber $\mathcal{F}(u)$ for the no-proximity effect model is very large while at the same time the data table $f$ has some small entries including a zero entry. These observations point to using the 
well-established standard Markov Chain Monte Carlo method
for running the goodness-of-fit test. One might think that computing a Markov basis for the no-proximity effect model (or equivalently the second hypersimplex $A(22)$)  is intractable. Luckily, 
a Markov basis (in fact a {\it Gr\"obner basis}) for this model 
is available, see \cite[Theorem 9.1]{GBCP} and \cite{DST}. Our Markov basis is defined by the following set of moves:   For each  $1 \leq i < j < k < \ell \leq 22$ one defines two moves $m[i,j; k,\ell]$ 
and $m[i,\ell;j,k]$. The first move is a table where the $(i,j)$ and $(k,\ell)$ entries are set to equal $1$,
the $(i,k)$ and $(j,\ell)$ entries are set to equal $-1$, and all the other entries are set to equal $0$. The second move is a table where the $(i,\ell)$ and $(j,k)$ 
entries are set to equal $1$, the $(i,k)$ and $(j,\ell)$ entries are set to equal $-1$, and all the other entries are set to equal $0$.
These tables  together with their negatives $-m[i,j;k,l]$ and $-m[i,\ell; j,k]$ comprise our  Markov basis $\mathcal{B}$. 
This Markov basis contains $ 2 \cdot 2 \cdot {22 \choose 4} \, = \, 29,260$ moves. 
\medskip

Using this MCMC we generated a set of $m$ random tables $f_1, \ldots, f_m$  using the Metropolis-Hastings algorithm and estimated the $p$-value of goodness-of-fit test 
by  $$ \frac{1}{m} \left( \sum_{\chi^2(f_i) \geq \chi^2(f)} 1\right).$$
One important parameter of the Metropolis-Hastings algorithm is the number of steps it requires between each selection of tables $f_i$ and $f_{i+1}$. The rule of thumb is that one needs sufficient 
number of steps  so that the Markov chain can reach any table in $\mathcal{F}(u)$ starting from an arbitrary table. We followed the method in \cite{Ido} to heuristically determine this number of steps:
It is a consequence of the Gr\"obner basis theory that there is a unique table $T_{unique}$  in ${\mathcal F}(u)$ where none of 
the Gr\"obner basis moves can be applied, and every table in ${\mathcal F}(u)$ is connected to $T_{unique}$ table via the Markov basis moves. This table is the unique reduced normal form of the data
table with respect to the Gr\"obner basis from which our Markov basis $\mathcal{B}$ is constructed. Empirically, the average number of steps one needs to go from a randomly generated table in $\mathcal{F}(u)$ 
to $T_{unique}$ using the moves in $\mathcal{B}$ is about $15000$. Hence, we estimate that the number of steps needed to connect two tables in $\mathcal{F}(u)$ is bounded by $30000$.

\subsection {Log-Ratio Test}
Since the single-pair proximity effect model contains the no-proximity effect model  we compared the relative fit of the two models by 
a \emph{ likelihood ratio test }.
The likelihood ratio test statistic is  defined as:
	\begin{equation} G^2=2\sum_{1 \leq j < k \leq 22} \hat f_{jk}^1 \log \left(\frac{\hat f_{jk}^1}{\hat f_{jk}^0}\right)
	\end{equation}
where  $\hat f_{jk}^1$ is the MLE  with respect to  the single-pair proximity effect model  and $\hat f_{jk}^0$ is the MLE with respect to the no-proximity effect model.  
It is well known \cite[Theorem 10.2.8]{LLM} that, for large sample sizes $N$, if the null hypothesis is true (i.e. if the data fits the no-proximity effect model better than it fits 
the single-pair proximity effect model) then $G^2$ has a $\chi^2$ distribution with 
degrees of freedom equal to the difference of the ranks of the nested log-linear models. In our case, the rank of the no-proximity effect model is equal to $\mathrm{rank} \, A(22) = 21$
and the rank of the single-pair proximity effect effect model is equal to $\mathrm{rank} \,  A[r,s] = 22$. 


\section{Numerical Results}

\begin{table}[h]
	\begin{center}
	\caption{ Maximum Likelihood Estimate  for experimental Table \ref{table:Lympho2004} obtained from irradiated  human lymphocytes.}
	{\tiny\tt
	\begin{tabular}{lllllllllllllllllllllll}

Chr&2&3&4&5&6&7&8&9&10&11&12&13&14&15&16&17&18&19&20&21&22&Sum\\
\hline1&47&43&38&37&33&33&27&34&26&28&25&23&22&24&23&20&14&8.9&15&8.8&11&541\\
2&&37&32&32&29&29&24&30&23&25&22&21&20&22&21&18&13&8.3&14&8.2&11&485\\
3&&&30&30&27&27&23&28&22&24&22&20&19&21&20&18&13&8.2&14&8.2&11&465\\
4&&&&27&24&24&21&26&20&22&20&18&18&19&18&16&12&7.8&13&7.8&10&425\\
5&&&&&24&24&21&26&20&22&20&19&18&19&19&17&12&8&13&8&10&426\\
6&&&&&&22&19&24&19&20&19&17&17&18&17&16&12&7.7&12&7.7&9.7&395\\
7&&&&&&&19&24&19&20&19&17&17&18&17&16&12&7.9&13&7.9&9.9&396\\
8&&&&&&&&20&16&18&16&15&15&16&15&14&11&7.2&11&7.2&9&345\\
9&&&&&&&&&20&21&20&18&18&19&18&17&13&8.6&13&8.5&11&416\\
10&&&&&&&&&&17&16&15&15&16&15&14&11&7.3&11&7.4&9.2&338\\
11&&&&&&&&&&&17&16&16&17&16&15&11&8&12&8&9.9&364\\
12&&&&&&&&&&&&15&15&16&15&14&11&7.6&12&7.6&9.4&337\\
13&&&&&&&&&&&&&14&15&15&13&10&7.4&11&7.4&9.1&319\\
14&&&&&&&&&&&&&&15&14&13&10&7.4&11&7.4&9&310\\
15&&&&&&&&&&&&&&&15&14&11&7.9&12&7.9&9.6&332\\
16&&&&&&&&&&&&&&&&14&11&7.9&11&7.9&9.5&321\\
17&&&&&&&&&&&&&&&&&10&7.4&11&7.4&8.9&291\\
18&&&&&&&&&&&&&&&&&&6&8.6&6&7.3&223\\
19&&&&&&&&&&&&&&&&&&&6.4&4.6&5.6&156\\
20&&&&&&&&&&&&&&&&&&&&6.6&7.8&240\\
21&&&&&&&&&&&&&&&&&&&&&5.7&156\\
22&&&&&&&&&&&&&&&&&&&&&&193

		\end{tabular}}
	\end{center}
 	 	 \label{MLE:Lympho2004}
\end{table}

We first computed the MLE tables for the no-proximity effect and the single-pair models as discussed in the previous section. This algorithm preserves the minimal sufficient statistic 
which guarantees that the MLE table belongs to the fiber of the data table. The MLE table for the no-proximity effect model is displayed above. Note that we display just two significant digits.
The table for the no-proximity effect reveals some marked differences with the experimentally observed table. Most notably, entries for each chromosome tended to be more 
homogeneous than those in the experiment. This is particularly true for small chromosomes 19 to 22. 

\medskip

Using the MCMC approach explained earlier we generated $3 \times 10^{10}$ tables which were sampled every $ 3 \times 10^4$ times to reduce the intrinsic correlation in the Markov Chain. 
We therefore obtained a sample size of $10^6$ tables. The chi-square statistic for each of the sampled table was computed. 
Interestingly, none of the $10^6$ tables generated for Table \ref{table:Lympho2004} had a test statistic smaller than the experimentally observed.  
Hence the no-proximity effect model could not be rejected.   This result is somewhat surprising but can be explained by 
estimating the number of tables that are contained in the ellipsoid defined by (\ref{chi-square}). We will give the details of this estimation in the Appendix. 
These results show that the data can be well fit by the no-proximity model and 
that the relative positions of chromosomes are random.





\begin{table}[h!]  
       \begin{adjustwidth}{-1.8cm}{}
	\begin{center}
 	\caption{Table of  deviations between observed and MLE counts. Each entry is the difference between  the observed counts and MLE for each pair of chromosomes $j$ and $k$ computed for Table \ref{table:Lympho2004}.} 

	{\tiny\tt
		\begin{tabular}{rrrrrrrrrrrrrrrrrrrrrr}
	
Chr&2&3&4&5&6&7&8&9&10&11&12&13&14&15&16&17&18&19&20&21&22\\
\hline
1&-3.2&-5.5&4.2&-8.2&-7.2&-3.8&-9&5&3.3&-2.9&-7&-8.1&-4&10&8.4&2.2&-2&5.1&6.8&0.2&16\\
2&&6.2&4.6&-0.1&1.2&-4.6&1.1&-0.9&-6.9&-0.9&7.6&8.2&-11&4.4&-13&5.9&-4.9&-1.3&-2.1&4.8&4.3\\
3&&&-9.4&0.8&4.7&-3.2&-1.8&-2.4&1.1&1.2&1.4&0.9&-1.3&-2.8&-0.9&3.4&-1.7&8.8&-2.8&3.8&-0.5\\
4&&&&-3.9&2.5&3.6&3.3&0.3&0.0&-8.7&-0.8&4.5&4.2&0.8&-2.4&1.7&-0.9&-1.8&-0.9&2.2&-3\\
5&&&&&-7.5&6.6&5.2&-0.6&3.9&8.2&5.1&6.4&-3&-0.4&-11&2.5&0.9&-1&2.9&-1.0&-6.2\\
6&&&&&&-4.4&2.8&-2.5&12&-7.2&11&0.6&-1.8&0.9&-3.4&-0.5&1.5&2.3&-3.5&0.3&-2.7\\
7&&&&&&&0.7&-3.5&-1.7&7.7&6.4&-4.5&1.1&-10&0.5&7.3&-0.7&1.1&6.3&-1.9&-2.9\\
8&&&&&&&&-7.2&-4.3&6.4&-5.2&9.7&0.1&0.0&-3.5&2&6.5&-3.2&-2.4&-0.2&-1\\
9&&&&&&&&&1.2&3.7&-13&4.5&5.1&7.9&1.6&-1.6&9.5&-0.6&-4.5&-1.5&-0.7\\
10&&&&&&&&&&0.8&5&-1.1&-0.7&-5.8&3.7&0.1&-1.6&-2.3&-0.4&-0.4&-6.2\\
11&&&&&&&&&&&7.8&-11&-0.9&-1.0&2.6&0.1&-3.4&2&-0.3&-5&1.1\\
12&&&&&&&&&&&&-6.2&1.2&-6.8&-3.3&2&-2.8&5.4&-1.6&-2.6&-4.4\\
13&&&&&&&&&&&&&15&-5&-4.6&-6.4&5.6&-2.4&-5.2&-0.4&-0.1\\
14&&&&&&&&&&&&&&7.3&-1.3&-7.1&-0.2&-5.4&-5&5.6&2\\
15&&&&&&&&&&&&&&&6.8&-0.9&-2&-0.9&-0.7&-0.9&-0.6\\
16&&&&&&&&&&&&&&&&-1.6&4.2&4.1&8.5&0.1&3.5\\
17&&&&&&&&&&&&&&&&&-5&-3.4&0.3&-2.4&1.1\\
18&&&&&&&&&&&&&&&&&&-4&2.4&3&-4.3\\
19&&&&&&&&&&&&&&&&&&&-0.4&-4.6&2.4\\
20&&&&&&&&&&&&&&&&&&&&0.4&2.2\\
21&&&&&&&&&&&&&&&&&&&&&0.4\\

		\end{tabular}}
	\end{center}

	 	 \label{table:Lympho2004Dev}
\end{adjustwidth}
\end{table}

\begin{table}
\begin{center}
	\caption{Table of pairs of chromosomes with their corresponding Chi-squared value and $p$-value corresponding to their extended interaction model and $p$-value adjusted with the Bonferroni correction. Significant $p$-values are labeled in boldface.}
	\begin{tabular}
	  {c | c| c| c|}
	  \cline{2-4}
& \multicolumn{3}{|c|}
	{ $p$-value }\\
	\hline
\textbf{\em Chromosome Pair }
	&  Chi-squared
	&  p-value before correction
	&  Bonferroni corrected p-value \\ 
	\hline \hline
	  
	$\set{1,22}$ &   {\textbf{17.27}}& {\textbf{0.00005}}& \textbf{0.00138} \\
	$\set{13,14}$& {\textbf{13.66}}& {\textbf{0.00022}}& \textbf{0.01012}  \\
	$\set{3,19}$& 7.87 & 0.00502 & 0.23092 \\ 
	$\set{6,10}$& 7.78 & 0.00527 & 0.24242 \\
	$\set{6,12} $&  6.85 & 0.00888 & 0.40848 \\
	$\set{9,18} $&6.51 & 0.0107 & 0.4922 \\
	$\set{16, 20} $& 5.73 & 0.01671 & 0.76866  \\
	$\set{8, 13} $& 5.72 & 0.01673 & 0.76958 \\
	\hline
	\end{tabular}

	\label{table:pVal}
\end{center}
\end{table}

\medskip
Microscopy observations however have shown that some groups of chromosomes tend to be close to each other and form exchange type aberrations more frequently than what one would predict using the 
no-proximity effect mode \cite{Branco2006}. These chromosomes include those that are found
 in the center of the nucleus \cite{Ars2004,Corn2002} and those that form the nucleolus \cite{Ars2004}. This observation is
further supported by the large positive deviations between entries in the experimentally observed table and the MLE table. 
The differences between the observed table and the MLE table are shown in Table  \ref{table:Lympho2004Dev}. 
A positive entry indicates that the observed table had more exchanges than the number predicted by the MLE table while a negative entry shows cases where the MLE table had more exchanges 
than the observed table.  The largest positive entries
in the table are those for chromosome pairs \{1,22\}, \{13,14\}, \{1,15\}, and \{9,18\}. We therefore tested all chromosome pairs on the single-pair proximity effect model and  performed the log-ratio tests considering the original log-linear models  against the modified modelsfor all chromosome pairs. 
We found eight pairs of chromosome pairs that were significant (Column 1 in Table \ref{table:pVal}).
However only two of them were significant after correction for multiple testing using Bonferroni  \cite{Bonf} (Column 4 in Table  \ref{table:pVal}).

\section{Discussion}

In this work we have presented a model-based approach to determine the relative positioning of chromosome territories from ETCA frequency 
tables  that are sparse and with small entries. 
In previous work a method for dealing with small entries was reported \cite{Corn2002},  however the assignment of p-values to specific clusters of 
chromosome territories was based on groups of chromosomes previously found in the literature. Our method 
builds on the techniques developed in classical algebraic statistics by Diaconis and Sturmfels \cite{Sturmfels98}. 

\medskip
Our results show that the overall distribution of chromosome exchanges can be simply explained by a model in which the relative position of chromosome territories is random. 
This finding does not quantitatively agree with the results reported in \cite{Ars2004,Corn2002} since a small deviation from randomness was reported in those studies. 
However all studies agree upon a rather random organization of chromosome territories. Several sources can be contributing to this apparently random organization of 
the genome. The first is imposed by the limitation of the data. Chromosome painting techniques are limited by the fact that homologue chromosomes are painted the same 
color, and this evidently introduces unavoidable noise since territories of homologous chromosomes can be positioned in very different environments (i.e. with different neighbouring chromosome territories). It is also possible that cell to cell variation is very large or that there is a severe reorganization of the chromosome territories after the radiation insult. The fact that new sequencing analysis is consistent with this overall picture suggest that radiation has a small repositioning effect.

Our study shows that a small fraction of chromosome pairs deviate from this picture 
of random ETCA formation. We identified eight pairs of chromosome territories that were significant prior to multiple testing correction (Table 4). 
The first two pairs of chromosome territories (ie.  $\{13,14\}$ and $\{1,22\}$) were also significant in \cite{Ars2004,Corn2002,Vives05}. There is an easy explanation for the significance  of these pairs although their true functional significance remains to be determined \cite{Bickmore02} . The pair $\{13,14\}$ is part of the cluster of chromosomes in the nucleolus $\{13,14,15,21,22\}$, an organelle that brings chromosome territories together for specific needs of the cell. 
The second pair  $\{1,22\}$ has been found to be part of a cluster of chromosome territories $ \{1,16,17,19,22\}$ located in the center of the nucleus of lymphocyte cells. We used a Bonferroni correction method for multiple testing. This method is known to be very conservative and it is possible that we rejected some informative pairs. In fact all the  pairs that were not significant after Bonferroni correction have been reported in several clinical blood malignancies suggesting that proximity of these pairs of chromosome territories may be somewhat common and furthermore may have an important 
role in the development of these diseases 
\cite{Bassat90,Bossi2014,Impera2011,Jarosova14,Naeem95,Zhu07}.

It is our intention to improve our results by including better outlier detection tools that help identify other chromosome pairs \cite{Rapallo, KRR} and by incorporating these results into the developent of  biophysical models. 
These models are based on different properties of the genome that can obtained from basic physical priciples such as the radial organization of chromosomes 
using overlapping sphere or ellipsoid packings \cite{Cremer96,Khalil,UW} , gene density \cite{Kreth04} or DNA decondensation processes \cite{Rosa08}  or through the folding of chromatin fibers \cite{Barbieri12,Blackstone11,Lieberman09}



\medskip
\textbf{Acknowledgements} This work is partially supported by NSF grant
DMS-1217324 and NIH grant RO1-GM109457 to J. Arsuaga. We want to thank J. L. Garc\'{i}a from Centro de Investigaci\'{o} del cancer de La Universidad de Salamanca (Spain) for sharing Figure 1.

\section{Appendix}
In this last section we return to our remark that none of the $10^6$ tables generated for Table \ref{table:Lympho2004} in our MCMC procedure had a test statistic smaller than $\chi^2(f)$.  In order 
to give a heuristic explanation for this behavior we will estimate the size of $\mathcal{F}(u)$ and also estimate the size of the set of the tables $F$ where $\chi^2(F) \leq \chi^2(f)$. 

\medskip We first give a lower bound for the cardinality of $\mathcal{F}(u)$, which is the set of lattice points in a polytope. A standard computational tool such as {\tt Latte} \cite{Latte} cannot compute
the cardinality of this immensely large set.  Instead we employ a divide-and-conquer approach where we consider the subtables consisting of the chromosomes 1 through 8 (Subtable A), 
chromosomes 8 through 15 (Subtable B), chromosomes 15 through 22 (Subtable C) in Table \ref{Subtables:Lympho2004}. 

\begin{table}[h]
     \caption{Subtables A-C}
	$\begin{array}{cc}
	{\tiny\tt
	\begin{tabular}{lllllllll}

Chr&2&3&4&5&6&7&8&Sum\\
\hline
1&44&38&42&29&26&29&18& 226\\
2&&43&37&32&30&24&25&235\\
3&&&21&31&32&24&21&210\\
4&&&&23&27&28&24&202\\
5&&&&&17&31&26&189\\
6&&&&&&18&22&172\\
7&&&&&&&20&174\\
8&&&&&&&&156\\
		\end{tabular}}  &
	{\tiny\tt
	\begin{tabular}{lllllllll}

Chr&9&10&11&12&13&14&15&Sum\\
\hline
8&13&12&24&11&25&15&16& 116\\
9&&21&25&7&23&23&27&139\\
10&&&18&21&14&14&10&110\\
11&&&&25&5&15&16&128\\
12&&&&&9&16&9&98\\
13&&&&&&29&10&115\\
14&&&&&&&22&134\\
15&&&&&&&&110\\
		\end{tabular}}  \\ 
	{\tiny\tt
	\begin{tabular}{lllllllll}

Chr&16&17&18&19&20&21&22&Sum\\
\hline
15&22&13&9&7&11&7&9& 78\\
16&&12&15&12&20&8&13&102\\
17&&&5&4&11&5&10&60\\
18&&&&2&11&9&3&54\\
19&&&&&6&0&8&39\\
20&&&&&&7&10&76\\
21&&&&&&&6&42\\
22&&&&&&&&59\\
		\end{tabular}}  &
	\end{array}$

	 	 \label{Subtables:Lympho2004}
\end{table}
\noindent
The $j$th entry in the {\it Sum} column in each subtable refers to the sum of $f(j,k)$ over all chromosomes $k\neq j$ included in the subtable. See for instance $1-8$ in Subtable A in Table \ref{Subtables:Lympho2004}. 
The remaining entries of Table \ref{table:Lympho2004} were subdivided into six rectangular subtables. Each of
these subtables are indexed by two subsets of chromosomes $J$ and $K$: Subtable 1 ($J= 1-4$, $K=9-15$),   Subtable 2 ($J= 5-7$, $K=9-15$),
Subtable 3 ($J = 1-4$, $K=16-22$), Subtable 4 ($J = 5-7$, $K=16-22$), Subtable 5 ($J = 8-11$, $K=16-22$), Subtable 6 ($J = 12-14$, $K=16-22$) in Table \ref{Subtables2:Lympho2004}.

\begin{table}[h]
	\caption{ Subtables 1-6}
	$\begin{array}{cc}
	{\tiny\tt
	\begin{tabular}{llllllll|l}

Chr&9&10&11&12&13&14&15&RSum\\
\hline
1&39&29&25&18&15&18&34 & 178\\
2&29&16&24&30&29&9&26  &  163\\
3&26&23&25&23&21&18&18  & 154 \\
4&26&20&13&19&23&22&20  & 143 \\
\hline
CSum & 120 & 80 & 87 & 90 & 88 & 67 & 98 \\
		\end{tabular}}  &
	{\tiny\tt
	\begin{tabular}{llllllll|l}

Chr&9&10&11&12&13&14&15&RSum\\
\hline
5&25&24&30&25&25&15&19  &  163\\
6&21&31&13&30&18&15&19  & 147\\
7&20&17&28&25&13&18&8    & 129\\
\hline
CSum & 66 & 72 & 71 & 80 & 56 & 48 & 46 \\
		\end{tabular}}  \\ 
\\
	{\tiny\tt
	\begin{tabular}{llllllll|l}

Chr&16&17&18&19&20&21&22&RSum\\
\hline
1&31&22&12&14&22&9&27 & 137\\
2&8&24&8&7&12&13&15  &  87\\
3&19&21&11&17&11&12&10  & 101 \\
4&16&18&11&6&12&10&7  & 80 \\
\hline
CSum & 74 & 85 & 42 & 44 & 57 & 44 & 59\\
		\end{tabular}}  &
	{\tiny\tt
	\begin{tabular}{llllllll|l}

Chr&16&17&18&19&20&21&22&RSum\\
\hline
5&8&19&13&7&16&7&4  &  74\\
6&14&15&13&10&9&8&7  & 76\\
7&18&23&11&9&19&6&7    & 93\\
\hline
CSum & 40 & 57 & 37 & 26 & 44 & 21 & 18 \\
		\end{tabular}}  \\ 
\\
	{\tiny\tt
	\begin{tabular}{llllllll|l}

Chr&16&17&18&19&20&21&22&RSum\\
\hline
8&12&16&17&4&9&7&8 & 73\\
9&20&15&22&8&9&7&10  &  91\\
10&19&14&9&5&11&7&3  & 68 \\
11&19&15&8&10&12&3&11  & 78 \\
\hline
CSum & 70 & 60 & 56 & 27 & 41 & 24 & 32 \\
		\end{tabular}}  &
	{\tiny\tt
	\begin{tabular}{llllllll|l}

Chr&16&17&18&19&20&21&22&RSum\\
\hline
12&12&16&8&13&10&5&5  &  69\\
13&10&7&16&5&6&7&9  & 60\\
14&13&6&10&2&6&13&11    & 61\\
\hline
CSum & 35 & 29 & 34 & 20 & 22 & 25 & 25 \\
		\end{tabular}}  \\ 
	\end{array}$

	 	 \label{Subtables2:Lympho2004}
\end{table}

\medskip
In the above tables, the $j$th entry in the {\tt RSum} column refers to the sum of the numbers in the $j$th row in the corresponding table, and 
the $k$th entry in the {\tt CSum} row refers to the sum of the numbers in the $k$th column in the corresponding table. Any table that has been subdivided 
in a total of $9$ subtables where subtables $A-C$ have same {\tt Sum} column as in the subtables $A-C$ of the data table and where 
 subtables $1-6$ have the same {\tt RSum} and {\tt CSum} columns/rows as in subtables $1-6$ of the data table is in $\mathcal{F}(u)$. So a lower bound for 
the cardinality of $\mathcal{F}(u)$ can be obtained by the product of the number of subtables of type $A,B,C$ and $1, \ldots, 6$ with the given {\tt Sum} and {\tt RSum}/{\tt CSum} columns/rows. 
The total number of such subtables range from $10^{34}$ for Table A to $10^{14}$ for Table \ref{Subtables2:Lympho2004}. The exact number of tables calculated by Latte \cite{Latte} are shown in 
Table \ref{LATTE:Lympho2004}.

\begin{table}[h]
\caption{ Total number subtables associated to subtables A-C and 1-6  }
$$ \begin{array}{c|c}
\mbox{Subtable} & \mbox{Size} \\
\hline
A  &  2952470953799239962752797659386190 \\
B &  252762217255461089482462934497 \\
C & 242451808378958740321921 \\
1 & 384937707376563538670706387547\\
2 &  11636397863410272633 \\
3 & 51895845228141509162048464\\
4 & 5538280355961059\\
5 & 336625602844011493310899\\
6  & 777971438252448\\
\end{array}
$$

	 	 \label{LATTE:Lympho2004}
\end{table}
The product of these numbers is in the order of $10^{214}$. 
This estimate however can be improved by the following arguments. Most of the $29,260$ Markov moves can be applied to Table \ref{table:Lympho2004} without changing the {\tt Sum} column. 
The few moves that cannot be applied are those that make the $(19,21)$ entry  negative. In addition, only $1554$ moves will not change the {\tt Sum} or {\tt RSum}/{\tt CSum} entries of the above 9 Subtables 
since the moves occur completely inside each of the subtables. 
The remaining $27,706$ Markov moves will alter at least two of these subtables so that their {\tt Sum} and/or {\tt RSum}/{\tt CSum} entries will change. If we repeat the above {\tt Latte} calculations 
for the nine subtables obtained after the application of each one of these $27,706$ Markov moves (in other words those that change {\tt Sum} and/or {\tt RSum}/{\tt CSum}) we will find that the number of these
new tables will be again of the order $6 \times 10^{214}$. We can repeat the same argument when two or more Markov moves are considered. For instance if we applied 
exactly 30 Markov moves of the 27706 possible 
moves in sequence then we would obtain a total number of new tables in ${\mathcal F}(u)$ given by $ 6 \times 10^{214}$ times
$${ 27706 \choose 30} \quad \approx \quad 7 \times 10^{100}.$$
This is approximately $4 \times 10^{315}$. 
This calculation however may have some tables that are counted more than once because two distinct sequences of moves starting from Table \ref{table:Lympho2004} can lead to
 the same 
table. This overcounting should be more than compensated by tables that could be reached with more or fewer than $30$ moves.
In conclusion we estimate that a lower bound of $ 4 \times 10^{315}$ possible tables is justified, but to be on the safe side   we adopt 
$10^{300}$ as a very conservative estimation of the  possible size of ${\mathcal F}(u)$.

\medskip
Now we provide a very liberal upper bound on the number of tables whose $\chi^2$ value is smaller than the $\chi^2$ value of the data table ($346.63$).  
The volume $V$ of the ellipsoid defined by  equation (\ref{chi-square}) is bounded above by $1.1 \times 10^{266}$ using the volume formula
for multidimensional ellipsoids.  It is known that 
the number of lattice points in an $n$-dimensional ellipsoid defined by $$\sum_{1 \leq j < k \leq n} a_{jk} x_{jk} \quad \leq \quad r^2$$ is approximately equal to $ V + O(r^{n/2})$ 
\cite{Landau} and  in our case $r^2 = 346.63$ and $n = 231$, and we arrive at $1.8 \times 10^{293}$ as an upper bound
on the number of tables which are inside the ellipsoid defined by (\ref{chi-square}).  We note that this must be a gross estimate, since we should be counting
the tables in the ellipsoid that are also in the fiber ${\mathcal F}(u)$. For this we should be counting the lattice points  in another ellipsoid of lower dimension $209$. 
We conclude that a very conservative estimate of proportion of tables $F$ where $\chi^2(F) \leq \chi^2(f) = 346.63$ in ${\mathcal F}(u)$ is extremely small:
$$ \leq \frac{1.8 \times 10^{293}}{10^{300}} \approx 1.8 \times 10^{-7}.$$ 
We believe that the true proportion is much smaller.

\end{document}